\documentclass{article}
\usepackage{spconf,amsmath,graphicx}
\usepackage{xcolor}
\usepackage{booktabs}
\usepackage{tabularx}
\usepackage{bm}
\usepackage[caption=false]{subfig}
\usepackage{flushend}
\usepackage[hidelinks]{hyperref}
\usepackage{extdash}

\usepackage{enumitem}
\setlist{nosep, leftmargin=14pt}

\usepackage{mwe} 

\title{Generation of anonymous chest radiographs using latent diffusion models for training thoracic abnormality classification systems}
\name{Kai Packhäuser, Lukas Folle, Florian Thamm, Andreas Maier
}
\address{Pattern Recognition Lab\\Department of Computer Science\\Friedrich-Alexander-Universität Erlangen-Nürnberg}
\begin{document}
%
\maketitle
\begin{abstract}
The availability of large-scale chest X-ray datasets is a requirement for developing well-performing deep learning-based algorithms in thoracic abnormality detection and classification. However, biometric identifiers in chest radiographs hinder the public sharing of such data for research purposes due to the risk of patient re-identification. To counteract this issue, synthetic data generation offers a solution for anonymizing medical images. This work employs a latent diffusion model to synthesize an anonymous chest X-ray dataset of high-quality class-conditional images. We propose a privacy-enhancing sampling strategy to ensure the non-transference of biometric information during the image generation process. The quality of the generated images and the feasibility of serving as exclusive training data are evaluated on a thoracic abnormality classification task. Compared to a real classifier, we achieve competitive results with a performance gap of only 3.5\,\% in the area under the receiver operating characteristic curve.
\end{abstract}
\begin{keywords}
Chest Radiography, Synthetic Image Generation, Abnormality Classification, Patient Privacy
\end{keywords}
\section{Introduction}
\label{sec:intro}
Over the past years, the availability of novel deep learning~(DL)~\cite{maier2019gentle} techniques has led to significant advances in the development of diagnostic algorithms for the detection and classification of thoracic abnormalities~\cite{guendel2018learning}. As the training of DL models is data-driven, it typically requires enormous amounts of training samples to obtain well-performing networks. However, medical images, e.\,g., chest radiographs, include personal information and are subject to certain privacy protection regulations, such as the EU General Data Protection Regulation (GDPR), which hinders the widespread sharing and, thus, utilization of acquired data. Therefore, to make large-scale medical image collections publicly available for research purposes, it is necessary to strictly comply with applicable objectives and requirements for the robust and reliable anonymization of sensitive patient data.

Conventional techniques for anonymizing chest radiographs rely on removing personally identifiable meta\Hyphdash information, e.\,g., patient names, or replacing them with pseudonyms. Furthermore, in the image domain itself, critical areas are typically obscured with black boxes. In one of our recent studies~\cite{packhauser2022deep}, however, we empirically demonstrate that such approaches do not offer sufficient means of protecting patient privacy. This is primarily because chest radiographs inherently contain biometric information (similar to a fingerprint) that can be exploited to successfully re-identify specific patients by DL-based linkage attacks.
Hence, there is an urgent need for more sophisticated privacy-enhancing techniques~(PETs) in medical imaging.

In this context, the generation of synthetic medical images with privacy guarantees has attracted attention as a promising solution for anonymization and overcoming data-sharing limitations. Generative models learn the probability distribution of a given real dataset and can be used to synthesize realistic representations by sampling from the learned data distribution. Provided that synthetic medical images are of sufficient quality, such data could be used to train DL models at scale. \mbox{Han et al.}~\cite{han2020breaking} utilized progressively growing generative adversarial networks~(PGGANs)~\cite{karras2018progressive} to synthesize chest radiographs and demonstrated the feasibility of GAN-based data augmentations. Moreover, they confirm the utility of exclusively synthetic training data in a clinical setting. Recently, diffusion probabilistic models~\cite{sohl2015deep} have become state-of-the-art for tasks such as high-resolution image synthesis. \mbox{Chambon et al.}~\cite{chambon2022adapting} leverage a latent diffusion model (LDM)~\cite{rombach2022high} for the generation of class-conditional chest radiographs. Nevertheless, the authors do not investigate whether the generated images can convey relevant class information during the training process of an abnormality classifier. Furthermore, their dataset is limited to only 2,000 samples.

In this work, we aim to investigate the feasibility of using exclusively synthetic training datasets for learning and recognizing thoracic abnormalities in chest radiographs. 
Therefore, we utilize an LDM~\cite{rombach2022high} to generate high-quality class-conditional images from a given real data distribution.
To synthesize a fully anonymous dataset from the trained generative model without transferring biometric patient identifiers, we propose a rigorous privacy-enhancing sampling strategy that excludes synthetic images, provided a specific patient identity has been learned from the original training data.
Finally, to quantify the quality of the generated anonymous dataset and to evaluate its applicability for training a thoracic abnormality identification system, we conduct multiple classification experiments and compare the performances of classifiers trained on real versus synthetic data, respectively.
Throughout our work, we leverage a \mbox{PGGAN}~\cite{karras2018progressive} as a synthetic image generation baseline.

\section{Methods}

\subsection{Dataset}
We use the large-scale ChestX-ray14 dataset~\cite{wang2017chestx} consisting of 112,120 chest radiographs from 30,805 individual patients. Accompanying metadata provides information about the corresponding 14 abnormality labels, including Atelectasis, Cardiomegaly, Consolidation, Edema, Effusion, Emphysema, Fibrosis, Hernia, Infiltration, Mass, Nodule, Pleural Thickening, Pneumonia, and Pneumothorax. Healthy subjects are labeled with an additional class, indicating that none of the above-mentioned abnormalities are present.
To simplify the class-conditional image generation task in this work, we exclude all images with multiple abnormality labels. Furthermore, we limit the maximum number of follow-up scans per patient to 5 to prevent the generative models from learning patient-specific patterns from over-represented subjects. Lastly, as the anatomy of young patients may not yet be fully developed, we only consider images from patients older than 21 years. After this data reduction procedure, the remaining 56,352 images are split into a training, validation, and test set by a ratio of 70{:}10{:}20. To this end, a patient-wise splitting strategy is applied to avoid potential patient overlap between the subsets while maintaining the overall class distribution.

\subsection{Latent diffusion model}
For the class-conditional image generation task, we apply the LDM proposed in~\cite{rombach2022high} leveraging pre-trained autoencoders. In this approach, input images are first embedded into a latent space of size 64$\times$64$\times$3 using the encoder of a vector quantized-variational autoencoder~(VQ-VAE) with 32, 64, and 128 channels in each stage. A diffusion model operates within that lower space dominated by lower frequencies and performs 1000 denoising steps with a U-Net (32, 128, and 256 channels). In the final step, the decoder of the autoencoder increases the spatial resolution to 256$\times$256 pixels while introducing higher frequencies. The class-conditional information is incorporated using a trainable lookup table. This is realized by combining the class embeddings with the diffusion process using cross-attention in the bottleneck of the U-Net.
We \mbox{closely followed the implementation by Rombach et al.~\cite{rombach2022high}}.

\begin{figure}
    \centering
    \includegraphics[width=\linewidth]{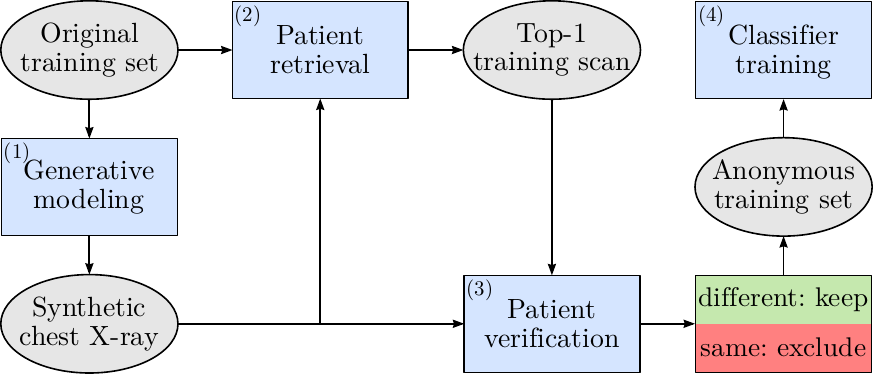}
    \caption{Illustration of the anonymous image generation process using the proposed privacy-enhancing sampling strategy. We apply generative modeling (1) techniques to synthesize chest radiographs. We then use patient retrieval (2) and verification (3) networks to ensure the non-transference of biometric identifiers. The resulting anonymous dataset maintains the size and class distribution of the original training set and is used for training an abnormality classifier (4).}
    \label{fig:sampling_strategy}
\end{figure}

\subsection{Progressively growing GAN}
As a synthetic image generation baseline, we employ the \mbox{PGGAN} architecture proposed in~\cite{karras2018progressive}. PGGANs are based on the idea of starting the training process with low-resolution images before gradually increasing the image resolution by adding layers to the network architecture. This allows the generator to focus on large-scale image structures at the beginning of the training procedure and to learn finer high-resolution details in later stages. We use a generator and discriminator that have the same general structure and grow equally over time. Both networks incorporate replicated \mbox{3-layer} blocks consisting of an upsampling or downsampling layer, a 3$\times$3 convolutional layer, and a leaky rectified linear unit (LReLU) with a slope of 0.2.
We adapt the originally proposed PGGAN architecture to produce chest radiographs with a final spatial resolution of 256$\times$256 pixels. The input to the generator is represented by a 256-dimensional latent noise vector. Its last 15 entries consist of the conditional vector, which controls the image class to be generated.
For the discriminator, we use a projection-based approach~\cite{miyato2018cgans} to incorporate conditional information into the network.

\subsection{Privacy-enhancing sampling strategy}
While synthetic data is often referred to as anonymous per se, generative models may memorize training examples and reproduce specific patient identities during inference~\cite{kuppa2021towards}.
To ensure the non-transference of biometric patterns when generating synthetic chest radiographs, we propose a rigorous privacy-enhancing sampling strategy~(see Fig.~\ref{fig:sampling_strategy}) that leverages state-of-the-art patient matching approaches. The strategy is based on the idea of excluding synthetic images in case patient-specific biometrics have been reproduced in a generated scan. To realize such an image selection process, we first apply a patient retrieval network for each generated chest radiograph to find its top-1 image in terms of patient similarity from the original training set. Then, we perform a 1-to-1 matching step for the recommended image pairs (real~vs.~synthetic) by applying a patient verification network that yields the probability of whether or not the synthetic image and the recommended real image belong to the same patient. Based on this probability value, we exclude all synthetic images that exceed a threshold of $t = 0.5$. We employ pre-trained patient retrieval and verification models proposed in previous work~\cite{packhauser2022deep}. The used networks achieve a top-1 precision of 99.6\,\% and an area under the receiver operating characteristic curve (AUC) of 99.4\,\%, respectively.

\subsection{Thoracic abnormality classification}
For the downstream abnormality classification task, we employ CheXNet~\cite{rajpurkar2017chexnet}, a densely connected convolutional network~(DenseNet)~\cite{huang2017densely} consisting of 121 layers. Its input layer receives chest radiographs with a resolution of 224$\times$224 \mbox{pixels}. The final classification layer yields a 14-dimensional output vector indicating the probability of the presence or absence of each abnormality class appearing in the \mbox{ChestX-ray14} dataset.

\section{Experiments}
All experiments were conducted using PyTorch and Python. The experimental setup included the following three steps.

\subsection{Training of the generative models}
The first step of our experimental pipeline consisted of training the generative models. For the LDM, we first trained the autoencoder with a combination of the perceptual loss~\cite{zhang2018unreasonable} and a patch-based adversarial objective~\cite{esser2021taming} for 100 epochs using a learning rate of $4.5 \cdot 10^{-6}$. Then, the autoencoder was applied to embed the input images into latent representations, which were used as inputs for the underlying diffusion model.
The diffusion model was trained with a learning rate of $10^{-6}$ using the $L_1$ loss until no further improvement in the validation loss (250 epochs).

The PGGAN architecture was trained using the Wasserstein loss with a gradient penalty coefficient of 10. For optimization, we used the Adam~\cite{kingma2014adam} optimizer with a learning rate of 0.001. The training was initiated at an image resolution of 4$\times$4. We gradually added layers to both the generator and the discriminator to increase the resolution by factors of~2 until an image size of 256$\times$256 was reached. Each stage of the \mbox{PGGAN} was trained until the generator and discriminator loss values stabilized~(100 epochs).

\subsection{Privacy-enhancing image generation}
After training the generative models, we synthesized two anonymous chest X-ray datasets by leveraging the trained LDM and PGGAN, respectively. For this purpose, we applied the proposed privacy-enhancing sampling strategy to ensure the non-transference of biometric identifiers during the image generation process. Note that the size and class distribution of the real training and validation set was maintained when creating the synthetic datasets. No oversampling of under-represented classes was performed. Both the LDM-based and the GAN-based datasets were used for further experiments as described in the following step.

\begin{table}[t]
    \centering
    \caption{Comparison of the abnormality identification performance of CheXNet when using either real or exclusively synthetic training sets. For all evaluations, the same real testing data is employed. We report the means and standard deviations of the resulting AUC values after 10 independent training and testing runs. Best synthetic results are shown in bold.}
    \resizebox{\linewidth}{!}{
    \newcolumntype{Y}{>{\centering\arraybackslash}X}
\newcolumntype{C}[1]{>{\centering\arraybackslash}p{#1}}
\begin{tabular}{lC{2.2cm}C{2.2cm}C{2.2cm}}\toprule
    Training set & Real & $\text{Syn}_{\text{PGGAN}}$ & $\text{Syn}_{\text{LDM}}$  \\\midrule
    Atelectasis         & $81.3 \pm 0.8$ & $70.1 \pm 1.2$ & $\bm{76.2 \pm 0.4}$ \\
    Cardiomegaly        & $92.9 \pm 0.6$ & $86.4 \pm 1.4$ & $\bm{88.6 \pm 0.8}$ \\
    Consolidation       & $74.8 \pm 1.0$ & $68.0 \pm 2.8$ & $\bm{75.5 \pm 1.1}$ \\
    Edema               & $92.8 \pm 0.8$ & $84.4 \pm 2.8$ & $\bm{87.5 \pm 1.7}$ \\
    Effusion            & $90.7 \pm 0.4$ & $83.2 \pm 0.9$ & $\bm{85.9 \pm 0.9}$ \\
    Emphysema           & $88.1 \pm 0.8$ & $76.5 \pm 1.6$ & $\bm{83.9 \pm 1.0}$ \\
    Fibrosis            & $80.8 \pm 1.0$ & $69.4 \pm 2.9$ & $\bm{77.3 \pm 0.6}$ \\
    Hernia              & $93.5 \pm 1.5$ & $80.6 \pm 3.5$ & $\bm{93.7 \pm 1.4}$ \\
    Infiltration        & $68.7 \pm 0.2$ & $59.1 \pm 0.6$ & $\bm{63.4 \pm 0.5}$ \\
    Mass                & $81.0 \pm 1.3$ & $67.7 \pm 0.9$ & $\bm{76.9 \pm 0.8}$ \\
    Nodule              & $71.4 \pm 1.0$ & $60.8 \pm 1.2$ & $\bm{67.8 \pm 0.7}$ \\
    Pleural Thickening  & $75.5 \pm 1.2$ & $68.4 \pm 1.3$ & $\bm{73.4 \pm 0.8}$ \\
    Pneumonia           & $70.8 \pm 3.3$ & $61.5 \pm 5.1$ & $\bm{65.8 \pm 2.0}$ \\
    Pneumothorax        & $79.9 \pm 0.7$ & $70.1 \pm 1.5$ & $\bm{77.0 \pm 0.5}$ \\\midrule
    Mean                & $81.6 \pm 0.4$ & $71.9 \pm 0.8$ & $\bm{78.1 \pm 0.3}$ \\\bottomrule
\end{tabular}

    }
    \label{tab:results_table}
\end{table}

\begin{figure*}[t]
    \hfill
    \begin{minipage}[b]{.49\textwidth}
    \centering
    \subfloat[Infiltration]{\includegraphics[width=.35\linewidth]{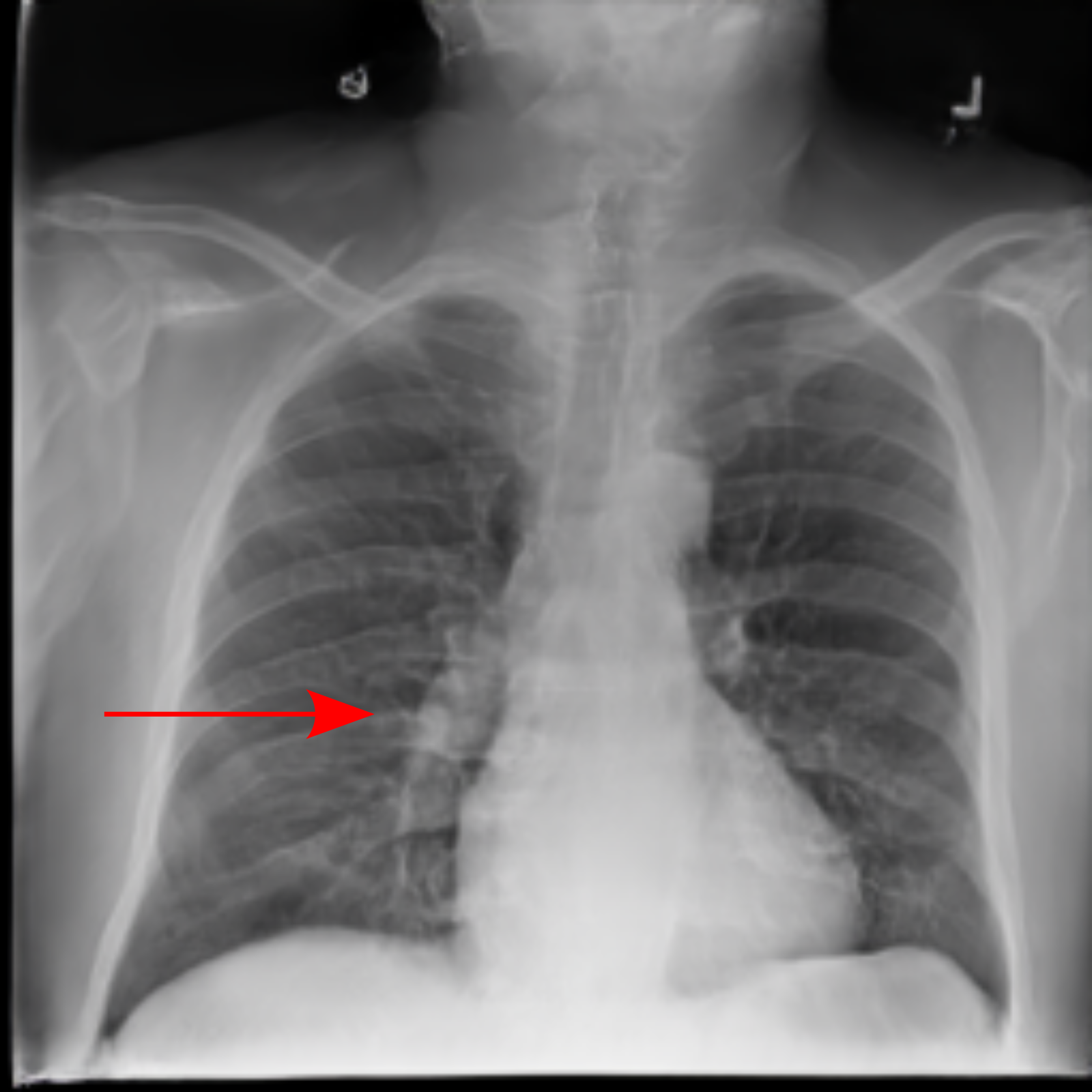}}
    \quad \quad \quad
    \subfloat[Nodule]{\includegraphics[width=.35\linewidth]{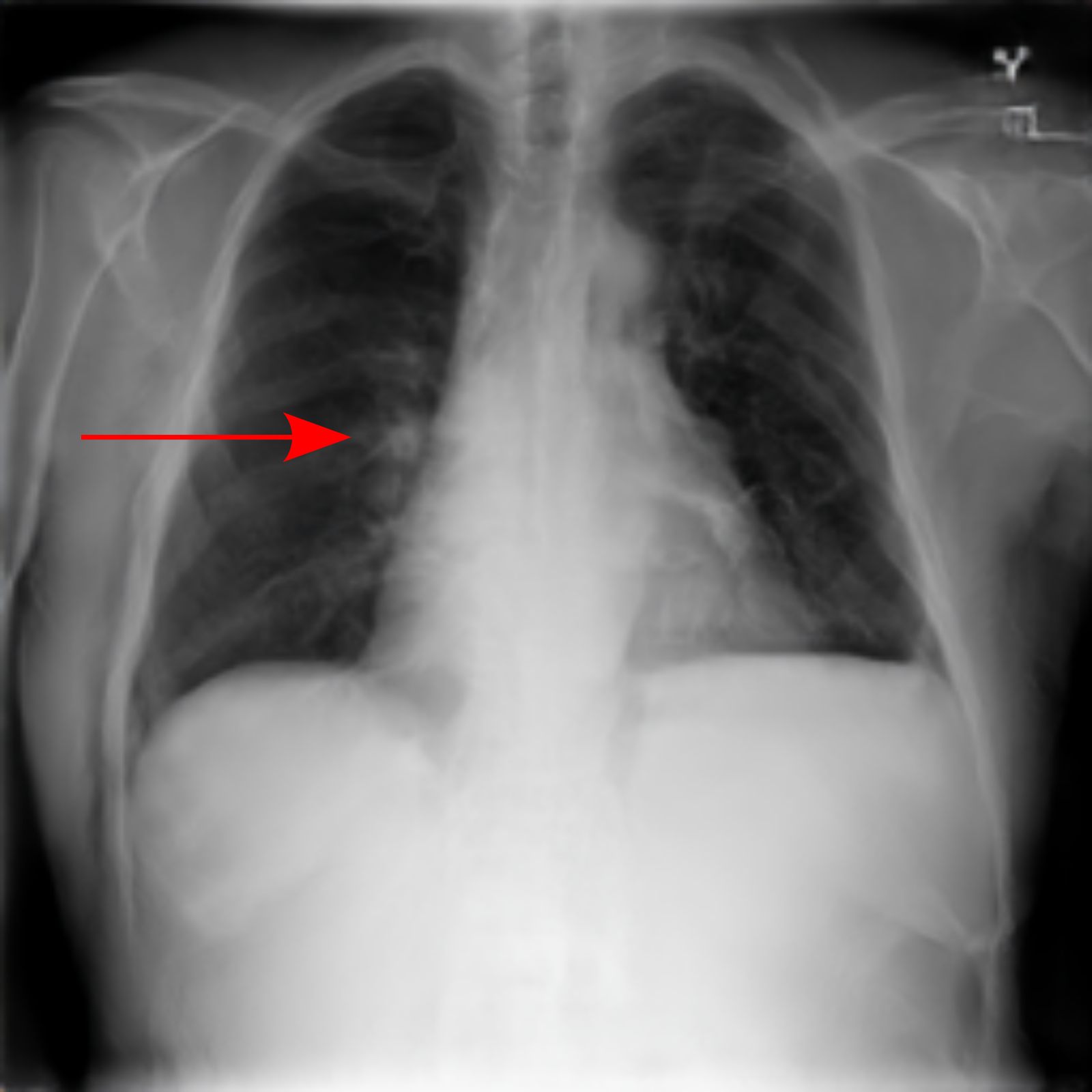}} \\
    \subfloat[Mass]{\includegraphics[width=.35\linewidth]{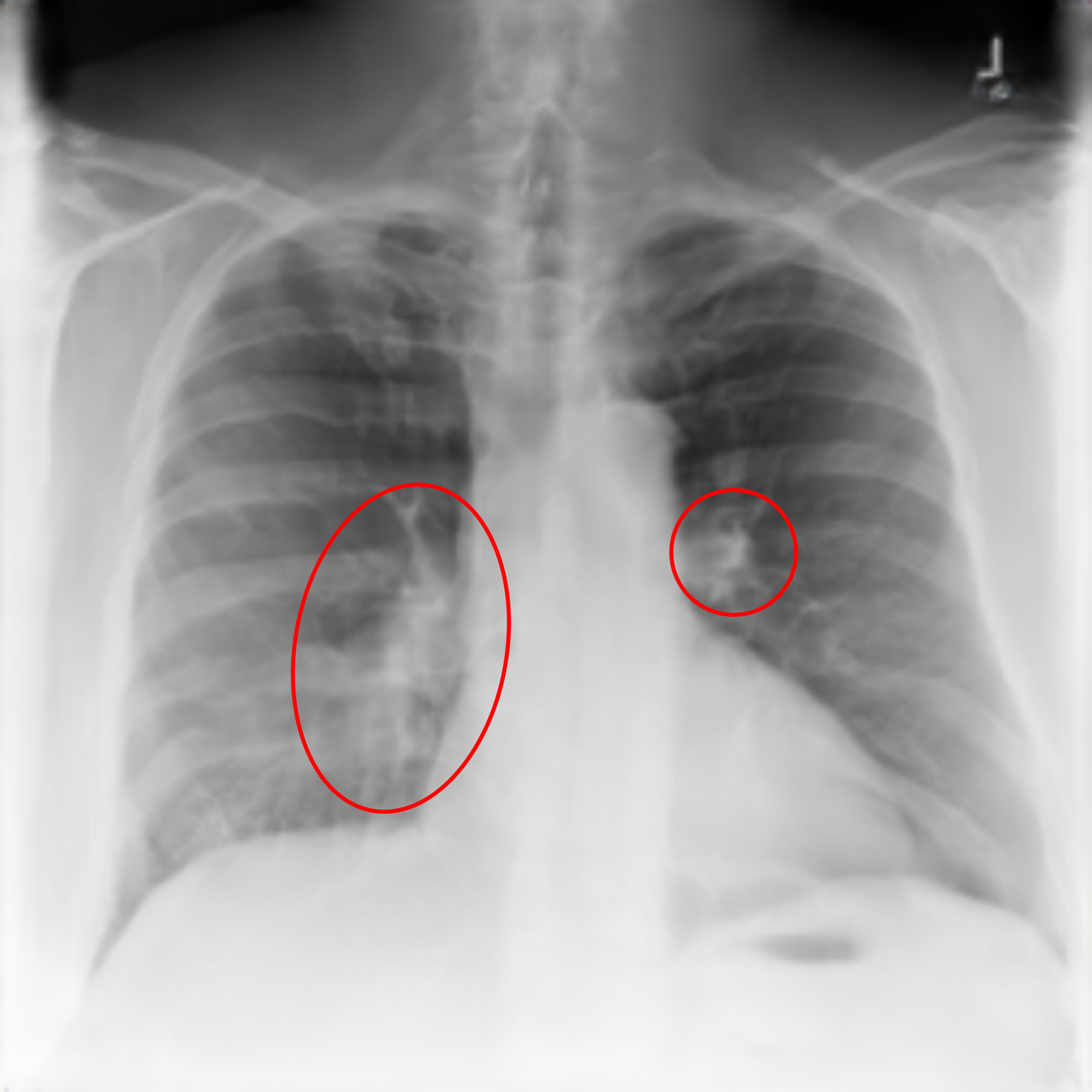}}
    \quad \quad \quad
    \subfloat[Cardiomegaly]{\includegraphics[width=.35\linewidth]{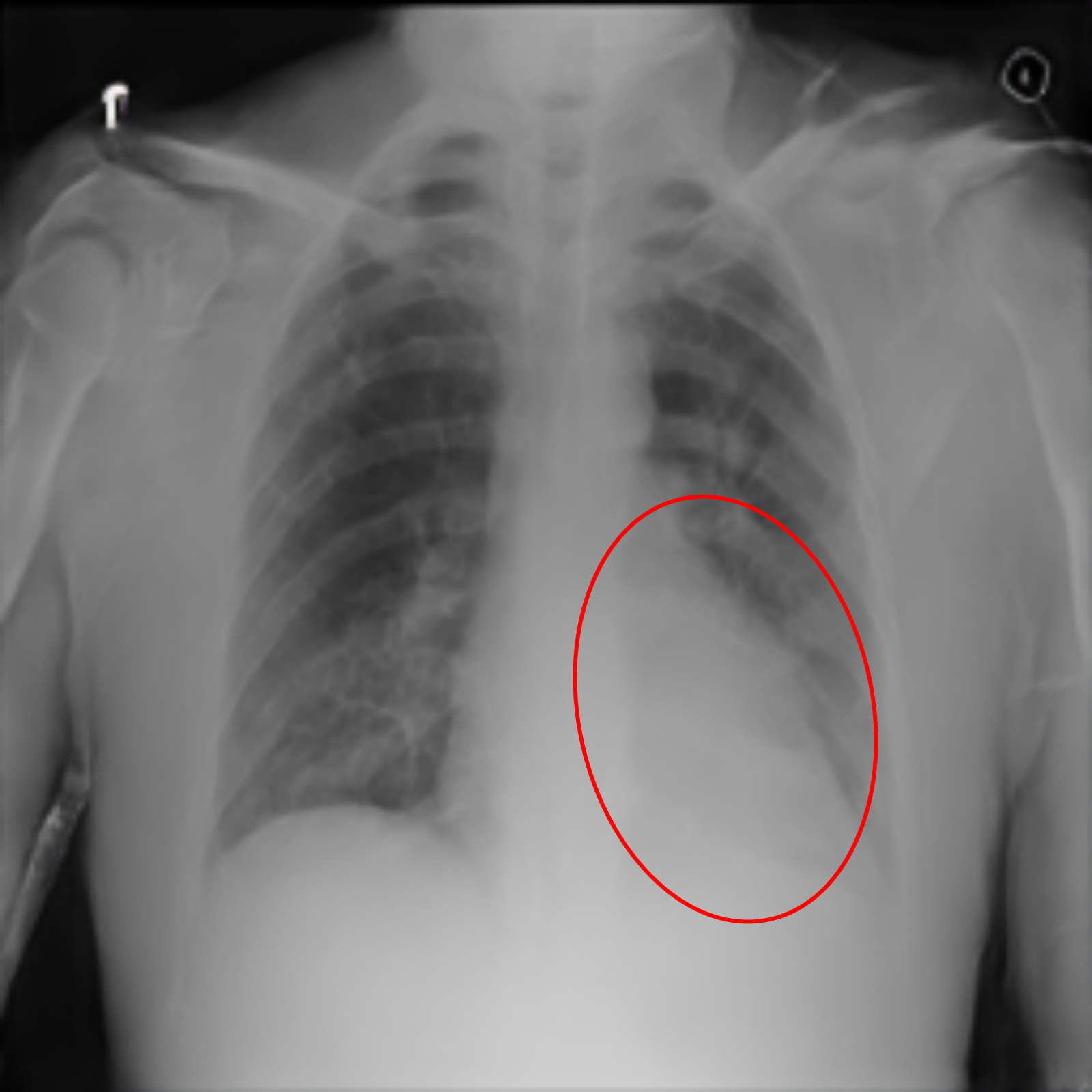}}\\[.5cm]
    \small(a) PGGAN
    \end{minipage}
    \hfill
    \begin{minipage}[b]{.49\textwidth}
    \centering
    \subfloat[Infiltration]{\includegraphics[width=.35\linewidth]{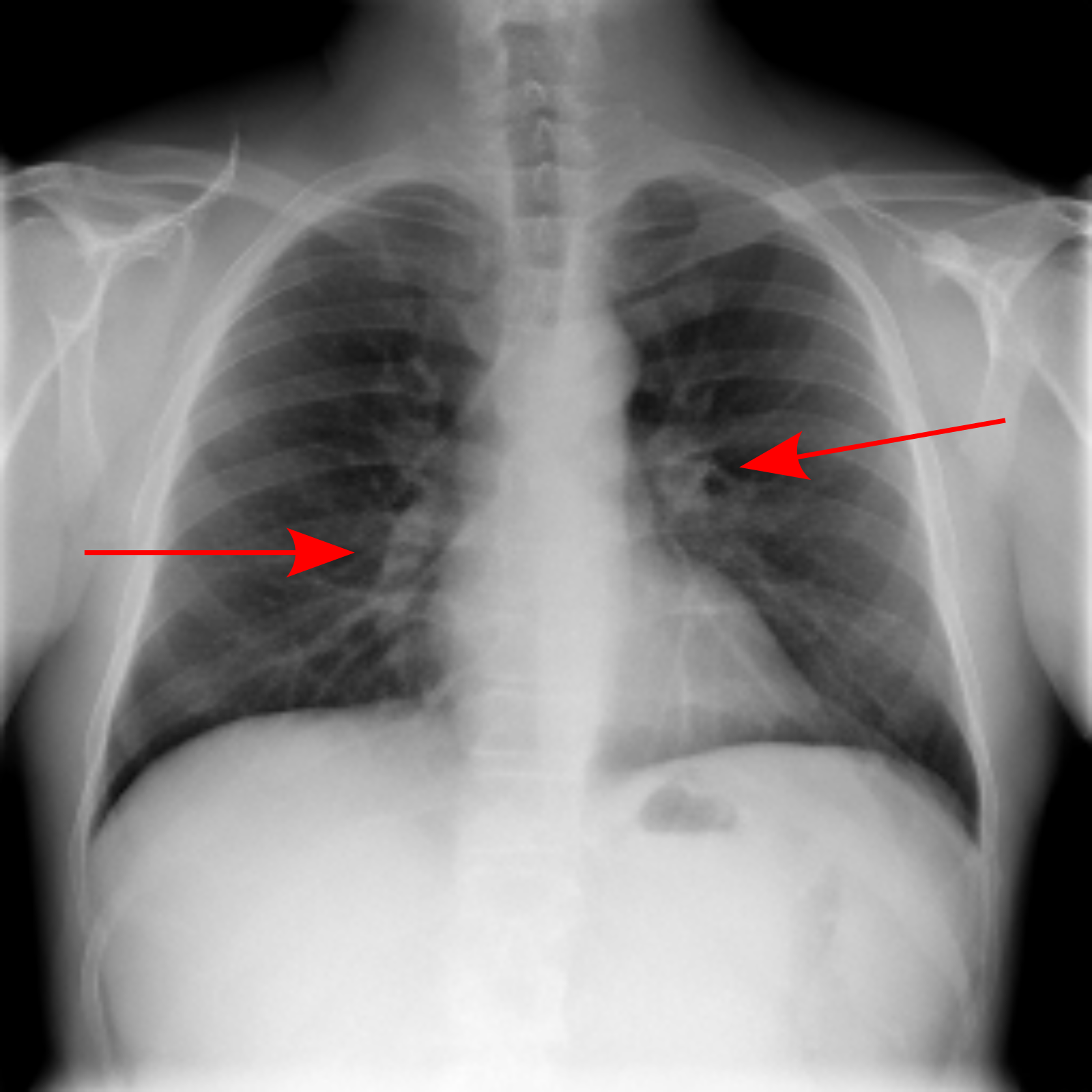}}
    \quad \quad \quad
    \subfloat[Nodule]{\includegraphics[width=.35\linewidth]{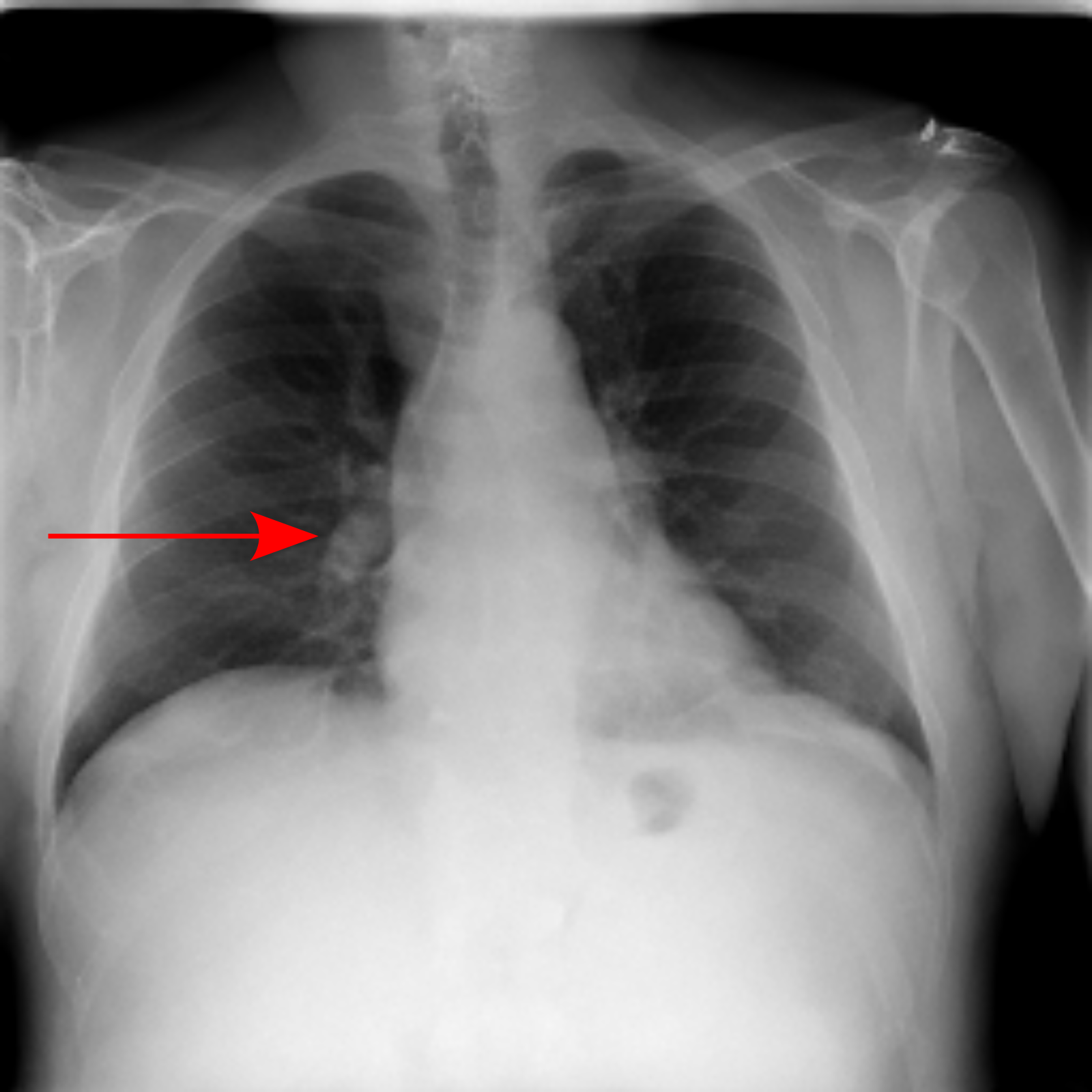}} \\
    \subfloat[Mass]{\includegraphics[width=.35\linewidth]{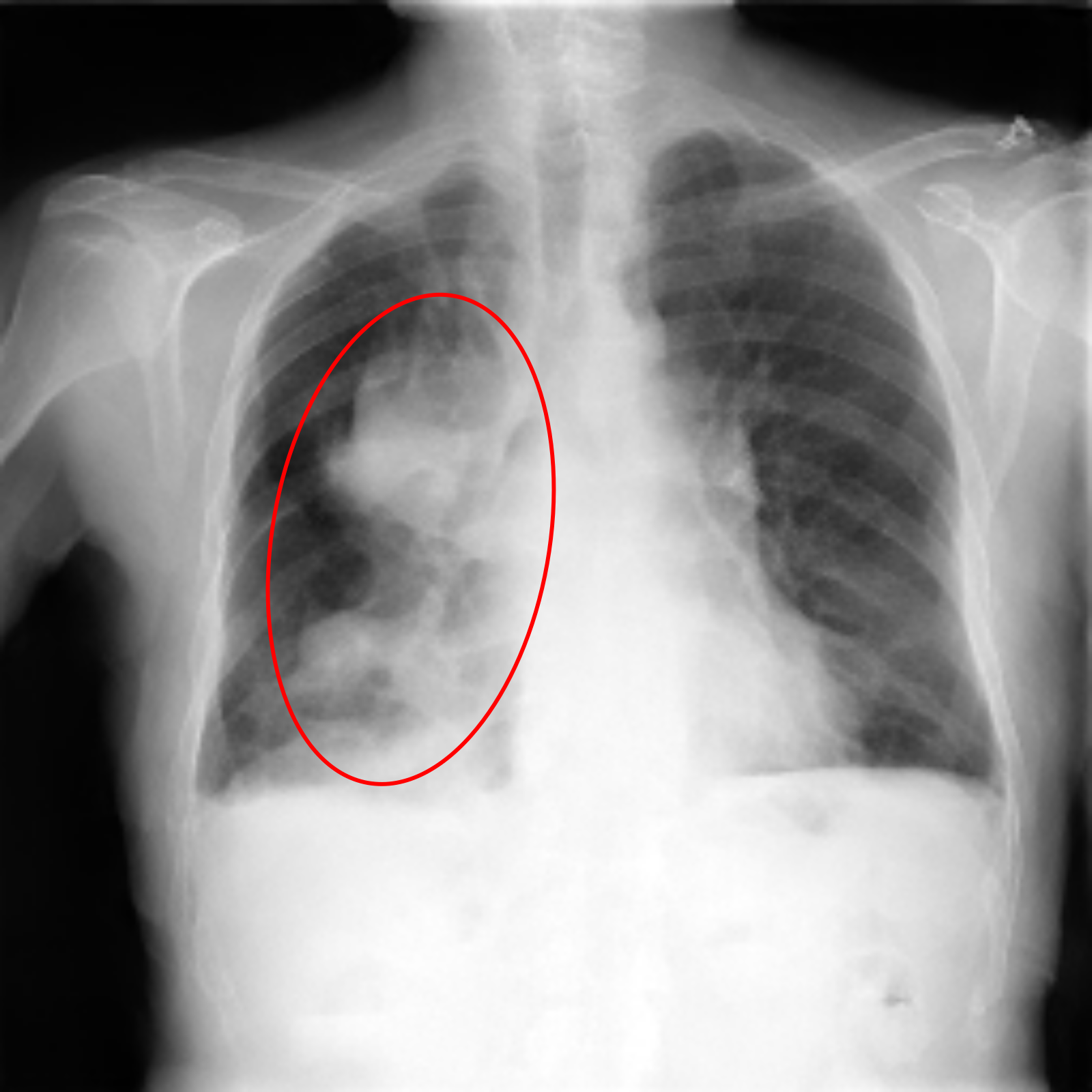}}
    \quad \quad \quad
    \subfloat[Cardiomegaly]{\includegraphics[width=.35\linewidth]{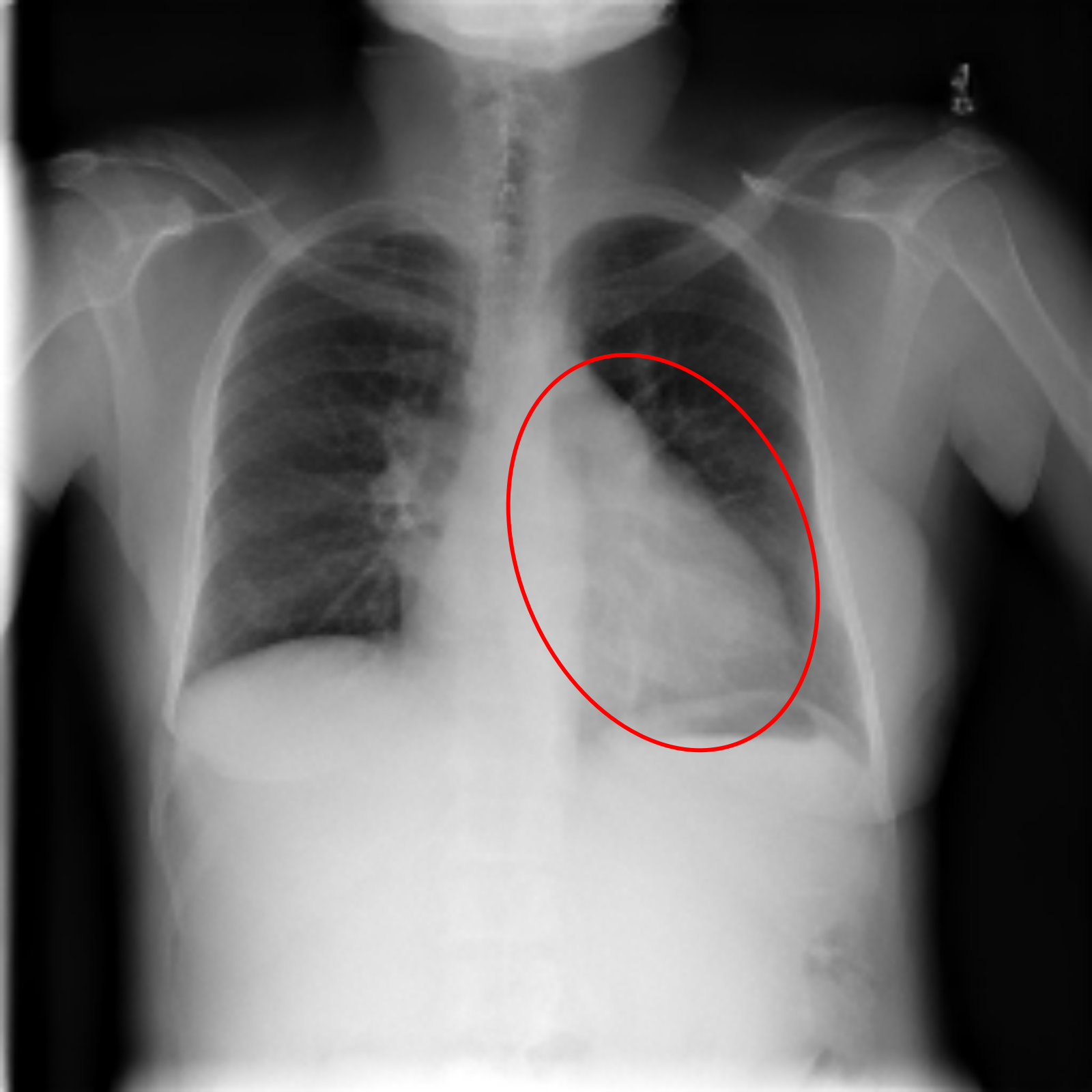}}\\[.5cm]
    \small(b) LDM
    \end{minipage}
    \hfill
    \caption{Randomly selected images generated by (a) the trained PGGAN and (b) the LDM, respectively. The positions and areas of induced abnormalities are highlighted with red arrows or circles.}\label{fig:images}
\end{figure*}

\subsection{Thoracic abnormality classification}
To assess the extent to which synthetic data can convey relevant class information during the training process of a thoracic abnormality classifier, we conducted experiments with both a real training set and with exclusive synthetic training sets generated by either the trained PGGAN or the LDM ($\text{Syn}_{\text{PGGAN}}$ and $\text{Syn}_{\text{LDM}}$). We trained the classifiers using the class-wise binary cross-entropy (BCE) loss and stochastic gradient descent (SGD) with a momentum of 0.9 and weight decay of 0.0001. The initial learning rate was set to 0.01 and was divided by a factor of 10 if the loss did not improve after an epoch. Model training was conducted until early stopping with a patience of 3. Each individually trained classifier was evaluated using the same real testing set. For each scenario, we performed 10 independent training and testing runs. The network performances are compared by reporting the means and standard deviations of the resulting AUC values.

\section{Results}
\subsection{Quantitative results}
The results of the conducted thoracic abnormality classification experiments are summarized in Table~\ref{tab:results_table}. As can be seen, the overall best performance is achieved when training the classifier with real images from the ChestX-ray14 dataset. In this case, we obtain a mean AUC of 81.6\,\%. Compared to this reference, the results significantly decrease for the classifier trained with the synthetic dataset generated by the \mbox{PGGAN}. The mean AUC of 71.9\,\% indicates that the GAN-based images fail to convey relevant class information during the training of the classification system.
In contrast, when using training images generated by the LDM, we observe competitive classification results with a mean AUC of 78.1\,\% on our test set.
Thus, the performance is only slightly below the reference, indicating the ability of the LDM-based chest radiographs to adequately preserve class-specific patterns. Interestingly, the performance for pathology classes Consolidation and Hernia is even higher than the reference.

\subsection{Qualitative results}
Randomly selected chest radiographs generated by the trained PGGAN and the LDM are shown in Fig.~\ref{fig:images}. In direct comparison, we perceive better image quality for the LDM-based chest radiographs. The trained LDM can produce high-resolution images with realistic patient anatomy. Moreover, as seen in the figure, each of the presented abnormalities becomes visible in the generated images.
In contrast, the GAN-based chest radiographs show a slightly worse image quality. This becomes apparent, for instance, by the partially unrealistic shape of the ribs and lungs, or through isolated artifacts that appear in the images. Moreover, we observe that the induced class conditions are a bit less pronounced in the images generated by the PGGAN.

\section{Discussion and Conclusion}
In this paper, we proposed a privacy-enhancing image sampling strategy to synthesize anonymous chest X-ray datasets. We leveraged an LDM and a PGGAN to generate high-quality class-conditional images. The obtained quantitative and qualitative results indicate the competitiveness of diffusion-based chest radiographs for training a thoracic abnormality classification system. With our work, we do not want to suggest that synthetic data can replace real data. In a real-life scenario, this would only be applicable if the quality of synthetic data is on par with real data. However, if this can be guaranteed, we hypothesize that our proposed approach allows for significant mitigation of data-sharing limitations.

In future work, next to further improving the overall image quality, 
we aim to exploit the generative models for compensating under-represented classes or creating infinitely large synthetic datasets. Lastly, we plan to incorporate a privacy-enhancing mechanism directly into the LDM to ensure the non-transference of biometric information during the training process.


\section{Compliance with ethical standards}
\label{sec:ethics}
This research study was conducted retrospectively using human subject data made available in open access by the National Institutes of Health (NIH) Clinical Center~\cite{wang2017chestx}. Ethical approval was not required as confirmed by the license attached with the open-access data.

\section{Acknowledgments}
\label{sec:acknowledgments}
The research leading to these results has received funding from the European Research Council (ERC) under the
European Union’s Horizon 2020 research and innovation program (ERC Grant no. 810316).
The authors gratefully acknowledge the scientific support and HPC resources provided by the Erlangen National High Performance Computing Center (NHR@FAU) of the Friedrich-Alexander-Universität Erlangen-Nürnberg (FAU). The hardware is funded by the German Research Foundation (DFG). The authors declare that they have no conflicts of interest.


\bibliographystyle{IEEEbib}
\bibliography{main}

\end{document}